\documentclass[twocolumn]{aa}
\usepackage{footnote, natbib}

\usepackage{color}

\usepackage{graphicx}
\usepackage{txfonts}
\usepackage[]{hyperref}

\newcommand{\anomalymatch}{\texttt{AnomalyMatch}}
\newcommand{\todo}[1]{\iffalse #1 \fi}
\begin{document}

\title{Identifying Astrophysical Anomalies in 99.6 Million Source Cutouts from the \textit{Hubble} Legacy Archive Using \anomalymatch{}}
\titlerunning{Identifying Astrophysical Anomalies Using \anomalymatch{}}

\author{David O'Ryan
      \inst{1}\thanks{Corresponding Author: david.oryan@esa.int}
      \and
      Pablo G\'omez\inst{1}
      }

\institute{$^{1}$European Space Agency (ESA), European Space Astronomy Centre (ESAC), Camino Bajo del Castillo s/n, 28692, Villaneuva de la Ca\~nada, Madrid
         }

\date{Received 14/05/2025; accepted 13/10/2025}
 
\abstract 
{}
{Astronomical archives contain vast quantities of unexplored data that potentially harbour rare and scientifically valuable cosmic phenomena. We leverage new semi-supervised methods to extract such objects from the \textit{Hubble} Legacy Archive.}
{We have systematically searched approximately 100 million image cutouts from the entire \textit{Hubble} Legacy Archive using the recently developed \anomalymatch{} method, which combines semi-supervised and active learning techniques for the efficient detection of astrophysical anomalies. This comprehensive search rapidly uncovered a multitude of astrophysical anomalies presented here that significantly expand the inventory of known rare objects.}
{Among our discoveries are 86 new candidate gravitational lenses, 18 jellyfish galaxies, and 417 mergers or interacting galaxies. The efficiency and accuracy of our iterative detection strategy allows us to trawl the complete archive within just 2–3 days, highlighting its potential for large-scale astronomical surveys.}
{We present a detailed overview of these newly identified objects, discuss their astrophysical significance, and demonstrate the considerable potential of \anomalymatch{} to efficiently explore extensive astronomical datasets, including, e.g., upcoming \textit{Euclid} data releases.}


\maketitle

\section{Introduction}\label{introduction}
\noindent With the launch of \textit{Euclid} \citep{2025A&A...697A...1E}, the beginning of operations of the Vera C. Rubin observatory \citep[][]{2019ApJ...873..111I} and the construction of the \textit{Nancy Grace Roman Space Telescope} \citep{2020JATIS...6d6001M}, the size of astronomical datasets is growing rapidly. These observatories will survey a large fraction of the sky in an agnostic way, leading to the potential discovery of a large number of new objects of astrophysical interest allowing us to expand our catalogues of rare galaxy morphology types, including cosmological effects like gravitational lenses or galaxies undergoing the effects of dense environments.

Rare objects, often termed astrophysical anomalies, are particularly informative for improving our understanding of galaxy evolution and cosmology. For example, strong lensing - a gravitational effect of chance alignment of galaxies - allows precise testing of the gravitational potential of the foreground galaxy, as well as the in depth study of the background galaxy from magnification effects \citep{2024SSRv..220...87S}. When observed around a galaxy cluster, such lensing is an excellent test of cosmology, allowing us to probe their dark matter halo. Other examples include jellyfish galaxies \citep{2016AJ....151...78P, 2017ApJ...844...48P, 2021A&A...648A..63D}, galaxy mergers \citep{2019A&A...626A..49P, 2018MNRAS.479..415A, 2024ApJS..274...23W}, edge-on protoplanetary disks \citep{2024ApJ...967L...3B} and Voorwerps \citep{2009MNRAS.399..129L}.

What many of these rare systems have in common are their methods of discovery. Often they are found by experts `manually' exploring their data, and detecting odd morphologies or serendipitously finding objects of interest \citep{2025A&A...696A.214P, 2025A&A...697A..14A}. Another common approach leverages the contribution of citizen scientists. Using platforms such as Zooniverse from the Galaxy Zoo collaboration \citep{2008MNRAS.389.1179L}, volunteers are able to mark any galaxies with anomalous morphologies as `odd'. 

However, these two searching strategies are time consuming and difficult to scale to large datasets. Additionally, they suffer particularly from the problem of subjectivity, and the difficulty in defining what an `odd' morphology is. To address this, the Galaxy Zoo: Weird and Wonderful project \citep{mantha20224} formalised the distinction between `interesting' and `non-interesting' anomalous galaxies in its classification scheme. Yet, volunteers may not be able to identify an anomaly as reliably as an expert in the field of extragalactic astrophysics. 

By their nature, anomalies are rare and difficult to find. Hence, applying commonly used machine learning techniques - such as supervised convolutional neural networks - is difficult due to limited training data. Samples containing the wanted anomaly are often small, and it is difficult to train models on limited training data for a highly imbalanced search space with mostly "uninteresting" data. 

Approaches such as \texttt{Astronomaly} \citep{2021A&C....3600481L} aim to resolve this problem. Rather than training to find individual types of anomalies, they utilise a combination of isolation forests \citep{liu2008} or local outlier factor \citep{Breunig2000} with active learning to identify anomalies specifically sought by a user. In further study, this can be combined with feature spaces produced by machine learning models \citep[e.g \texttt{Zoobot};][]{2022MNRAS.513.1581W, 2023JOSS....8.5312W} to identify further examples of an anomaly. This is effectively done with the \texttt{ulisse} tool \citep{2022A&A...666A.171D}, where these feature spaces can be analysed such that objects of similar morphologies appear close together and can be extracted from areas of interest.

Related works in the area often use unsupervised learning to search for astrophysical anomalies. Unsupervised methods in this context leverage unlabelled data and its data structure to create groups of similar objects expecting anomalies to cluster or fall outside clusters. Such an approach was successfully applied in \citet{2022ApJ...932..107S}. In this work, a self-supervised machine learning method was used to mine the entire Dark Energy Spectroscopic Instrument survey Data Release 9 for strong gravitational lenses. Their self-supervised method created an output representation space which was linked to the morphology of the objects trained upon. Once this feature space was created, the representation vector of a single image of a gravitational lens (output by the same model) was then used to conduct a similarity search through 76 million images, identifying 1,192 candidate gravitational lenses with excellent computational efficiency.

In this work we utilise a different approach. We frame anomaly detection as an imbalanced binary classification problem. This is not a fundamentally different task as anomaly detection inherently involves distinguishing rare instances from common ones, but it allows us to leverage semi-supervised learning techniques specifically designed for extreme class imbalance, enabling effective learning from minimal labelled anomaly examples. Our method utilises semi-supervised learning (SSL) techniques such as pseudo-labelling \citep{lee2013} and consistency regularization \citep{NIPS2014_8844a0d3} to make binary classifications on an anomalous or nominal source. Additionally, we use active learning, where experts iteratively validate and label predictions of the top ranking scored images. This hybrid approach offers key advantages: it requires very few initial labels (even fewer than ten anomalies), enables expert validation throughout the process, and effectively utilises both unlabelled data and expert knowledge to progressively improve detection performance. 

Our approach, named \anomalymatch{}, is fully described in the companion paper \citep{AMPaper}. There, we conduct thorough benchmarking and validation of the approach on established datasets such as mini-ImageNet \citep{Vinyals2016MatchingNetworks} and Galaxy MNIST \citep{walmsley2022galaxy}. In parallel to the development, we actively applied this approach to searching for anomalies in the entire \textit{Hubble Space Telescope} Legacy Archive\footnote{HLA: \url{https://hla.stsci.edu/}} (HLA).

We use the European Space Agency's (ESA) science data platform ESA Datalabs\footnote{ESA Datalabs: \url{https://datalabs.esa.int/}} \citep{2024sdm..book....1N} to conduct this search. In initial testing, we attempted to identify edge-on protoplanetary disks. These systems form very distinct `hamburger' or `butterfly' shapes which can easily be detected by image classification algorithms. However, only very few of them are known in the literature. As we developed and updated the algorithm, we expanded our search via active learning and marking other objects of astrophysical interest. This was due to multiple factors. First, was a limitation of the data we are using. We search primarily in a previously created dataset of $F814W$ filter \textit{HST} sources observed with the Advanced Camera for Surveys/Wide Field Channel (ACS/WFC; hereafter only ACS). In this dataset, the number of edge-on protoplanetary disks is limited as they are often observed at other wavelengths, in the Wide-Field Planetary Camera-2 (WFPC2) or are very difficult to resolve in \textit{HST} data \citep[][e.g. see]{1996ApJ...473..437B, 1998ApJ...501..841K, 2003ApJ...589..410S, 2008AJ....136.2136R, 2023ApJ...945..130A}. The second was that our algorithm began to detect numerous other objects of interest that we wished to pursue. These included lensed quasars, mergers and gravitational lenses.

In this work, we discuss and display the wealth of new objects found in the development and application of \anomalymatch{} and discuss the implications for our model. Overall, we report 1,176 newly found anomalies which span 19 different classes. These include galaxy mergers, gravitational lenses and arcs, edge-on protoplanetary disks, and a host of rare galaxy morphologies.

This paper is laid out as follows. Section \ref{data} describes the imaging data we use from the HLA, its creation, its potential limitations and the resultant training set we use. We briefly describe \anomalymatch{} in Section \ref{methods}, but primarily focus on our methods of anomaly extraction from the HLA and of searching the literature. We also briefly describe ESA Datalabs, and the efficiency of our model searching through our dataset. Section \ref{results} and \ref{discussion} show the anomalies we find through the HLA, and we discuss their representation in the literature and the new objects that we have discovered. Finally, Section \ref{conclusion} concludes this paper and we describe our plan for future implementation of \anomalymatch{}.

For an excellent description of an anomaly, see the introduction of \citet{2020arXiv200911732R}. In this work, we define an astrophysical anomaly (or simply anomaly) to be an astrophysical object that shows morphological characteristics which deviate notably from the general population.

\noindent Our contributions are as follows:
\begin{itemize}
    \item We conduct the first comprehensive systematic anomaly search of the entire HLA, comprising approximately 100 million image cutouts, using the recently developed semi-supervised and active learning method \anomalymatch{}.
    \item We present a substantial catalogue of newly identified astrophysical anomalies, significantly expanding the known populations of rare cosmic phenomena: 417 previously unknown galaxy mergers, 138 candidate gravitational lenses, 18 jellyfish galaxies, and 2 collisional ring galaxies.
    \item We demonstrate the exceptional efficiency and accuracy of our approach, processing the entire HLA dataset within just 2–3 days, highlighting its strong potential for rapid anomaly detection in upcoming large-scale astronomical surveys such as \textit{Euclid}.
\end{itemize}

\section{Data and training set} \label{data}
In this work, we use the source cutouts created in \citet{2023ApJ...948...40O}. In that work, we searched the HLA for interacting and merging galaxies, but created cutouts of every extended source within the archive. In this section, we briefly describe the creation process and comment on its limitations in finding other anomalous objects in the HLA. 

\subsection{Source cutouts of the HLA}\label{sec:or_cat}
\noindent The HLA contains all \textit{Hubble Space Telescope} (\textit{HST}) observations. These observations were obtained across many different dates, instruments and, filters. Therefore, to create a consistent dataset, \citet{2023ApJ...948...40O} elected to only use observations from the Advanced Camera for Surveys (ACS) Wide Field Channel with the $F814W$ filter. They also only used observations at Calibration Level 3 - i.e., science-ready mosaics from the general reduction pipeline. The source of these was the \textit{Hubble} Advanced Product dataset. This subsection of the HLA specifically contains curated \textit{HST} science-ready mosaics from across the sky.

With these selections, $\approx$10k observations were available. While these could be queried from well known Table Access Protocol services via a host like the Mikulski Archive for Space Telescopes this would lead to downloading terabytes of data. Therefore, the science platform ESA Datalabs was employed. This platform provides a direct link to various science archives and a Jupyter Notebook environment to give users direct access to \textit{HST} observations. The \textit{HST} data are stored in specific data volumes dependent on the instrument of observation. The Level 3, $F814W$ mosaics can be easily accessed with no downloading required.

Rather than conducting computationally expensive source extraction across such a large set of observations, \citet{2023ApJ...948...40O} employed the \textit{Hubble} Source Catalogue \citep[HSC;][]{2016AJ....151..134W}. This is a large, publicly available catalogue. \citet{2016AJ....151..134W} applied \texttt{SourceExtractor} \citep{1996A&AS..117..393B} across each observation in the HLA. Selecting only extended sources provided 99.6 million source cutouts in the image dataset.

For each cutout, a fixed size of 150$\times$150 pixels was used (7.5"$\times$7.5"), with a \texttt{LinearStretch} and \texttt{ZScaleInterval} from Astropy \citep{astropy:2013,astropy:2018,astropy:2022}. Each image was stored as a one-channel greyscale JPEG image. While this does lead to a loss of quality of the image, it allows for more efficient storage. The image creation process was prioritised such that the morphological features of each galaxy were visible. This approach, however, was not optimised to find low surface brightness features, potentially impacting our detection of anomalies such as jellyfish galaxies.

Upon the creation of the dataset in \citet{2023ApJ...948...40O}, it was found that many extended sources were shredded. This could result in an extended source appearing in the dataset multiple times, but from different centring points. The number of duplicates of extended sources depended on the source size and, therefore, the redshift of the source. The largest sources, we found, could be duplicated up to five times. These large sources would often be large enough to fill the $150 \times 150$ pixel source cutouts, and therefore, would show little detail. These were then given a low anomaly score by \anomalymatch{}.

Smaller sources, from $z > 0.2$, were less affected by shredding and duplication. In a minority of cases, these could be shredded up to three times. \citet{2016AJ....151..134W} did provide a \texttt{MatchID}, which attempted de-duplication of the HSC, but \citet{2023ApJ...948...40O} opted to conduct their own de-duplication after classification. This was motivated by preventing the removal of multiple galaxies that were involved in interactions and preserving the identification of merging galaxies with multiple cores. We conduct de-duplication of the data after the identification of anomalies. When we create our training set, we ensure there are no duplicate sources, or multiple shredded cutouts, of the source.

While only extended sources were selected in the HSC, many dense star fields - such as those in deep ACS observations of Andromeda and the Magellanic Clouds or globular clusters - were also flagged as extended. This occurred because the individual point sources within these regions were so densely packed that they were blended into a single extended source. This is a specific example of an image artifact which would make other methods of anomaly detection difficult. We identify this in the active learning of training our model, and ensure that \anomalymatch{} gives these a low anomaly score.

For ease of storage, the 99.6 million cutouts were are stored in $\approx1,000$ HDF5 files, containing $\approx100,000$ images each. By using this storage format, the images can be accessed and read to the memory efficiently.

\subsection{Training set}
\begin{figure}
    \centering
    \includegraphics[width=0.95\columnwidth]{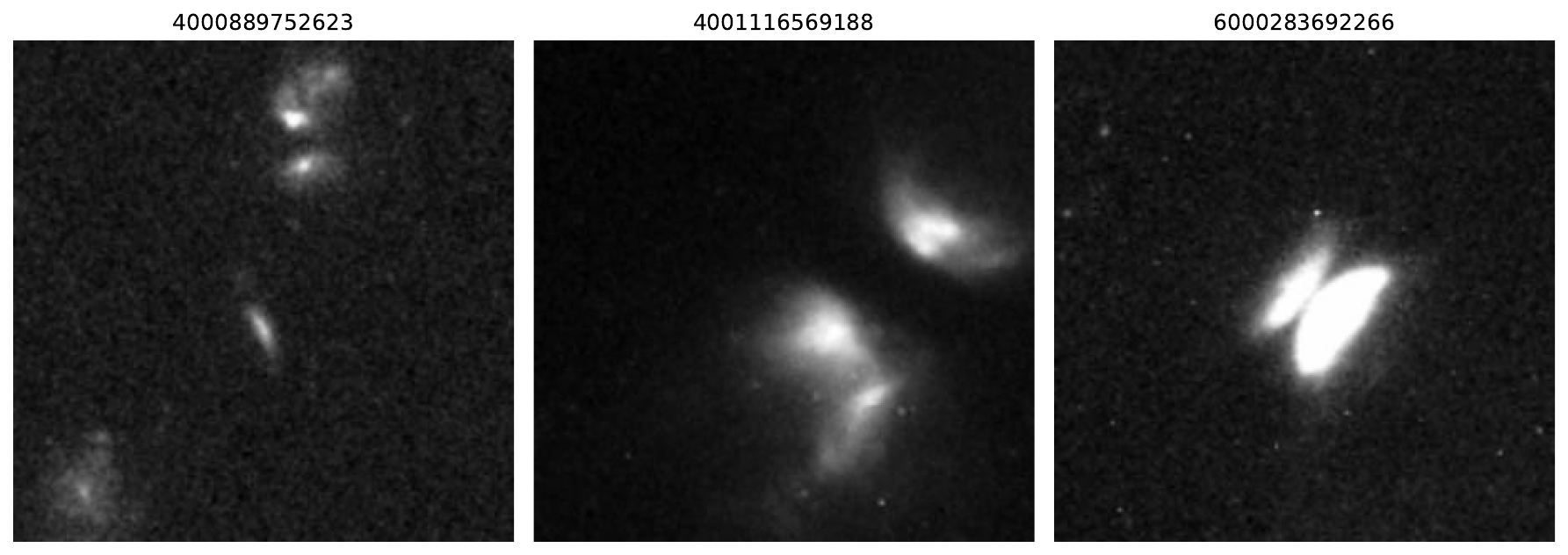}
    \caption{The initial three images labelled anomaly used to train \anomalymatch{}. These three images containing edge-on protoplanetary disks which we initially aimed to search additional instances of in the HLA. During active learning, this set was expanded to include sources with odd morphologies like mergers, lenses and jellyfish galaxies which were serendipitously discovered. Titles are the Source IDs of the objects found in \citet{2023ApJ...948...40O}.}
    \label{fig:proto-planetary_disks}
\end{figure}

Our initial goal was to find more examples of edge-on protoplanetary disks. Therefore, we started our search with three examples of these anomalies, a set 128 labelled nominal data and $\approx$99.6 million unlabelled images. The  nominal imaging was selected by visual inspection of randomly selected images from the full dataset. We selected images of individual galaxies, star fields and any imaging artifacts that appeared in the sample. Figure \ref{fig:proto-planetary_disks} shows the three anomalies we initially used to train \anomalymatch{}. However, upon conducting active learning, we found that the odd morphology of these systems led to the discovery of other objects of interest also being given high anomaly scores. With these new objects, we began to pivot our anomaly search to other interesting objects which were also more likely to appear in $F814W$ data. 

\begin{figure*}
    \centering
    \includegraphics[width=0.625\textwidth]{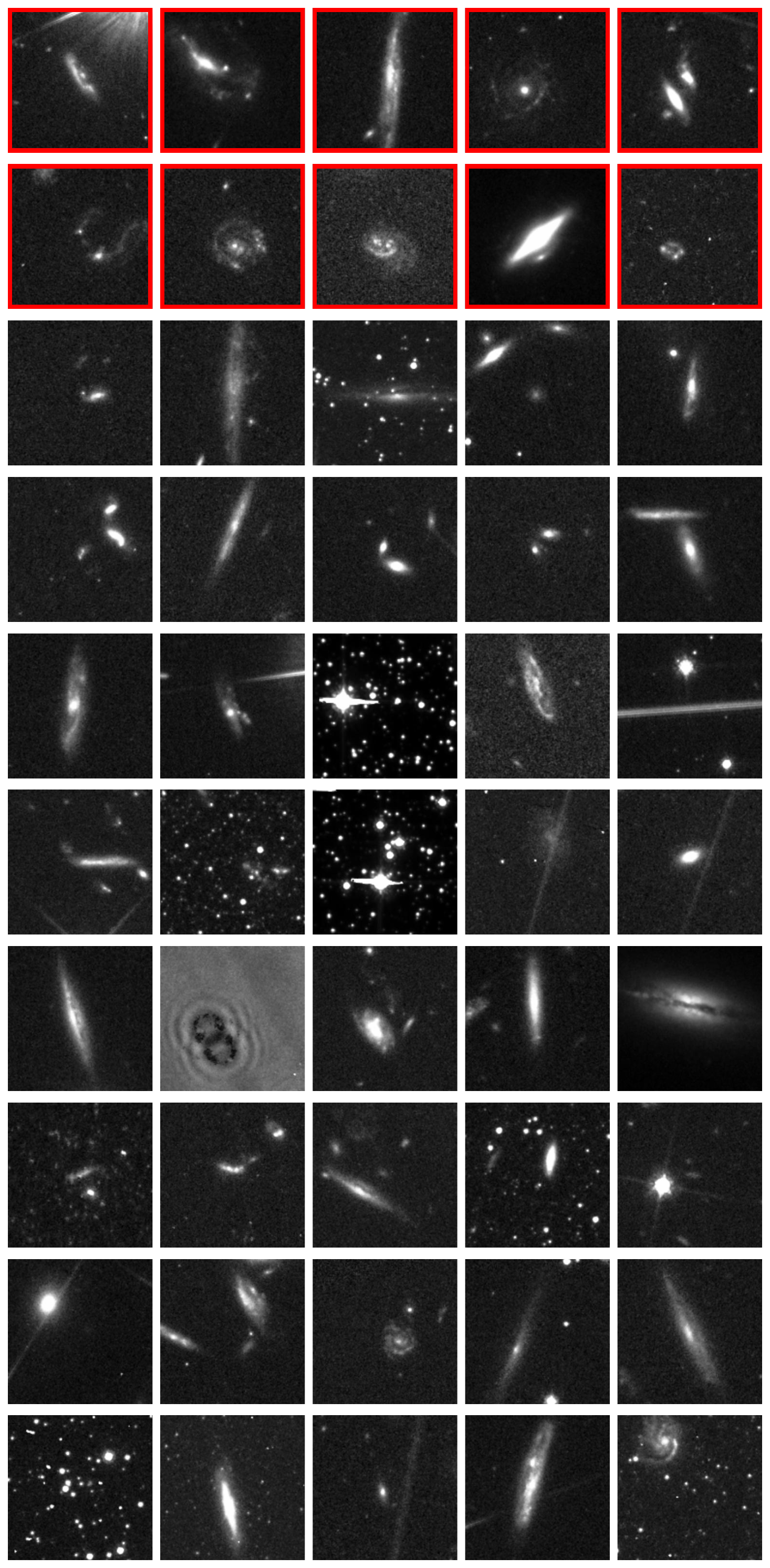}
    \caption{Fifty examples of the final training set used in applying $\anomalymatch$ to the HLA. The top two rows, highlighted in red, show ten examples of the anomaly class. These are made up of mergers, lenses, edge-on proto-planetary disks as well as some galaxies showing odd morphology. The remaining 40 images are then examples of `nominal' data. This is primarily isolated galaxies, star fields and artifacts.}
    \label{fig:final-training-set}
\end{figure*}

From this attempt to search for edge-on protoplanetary disks, we had constructed a second catalogue of interesting objects through active learning. The anomalies at this stage were classified based on their morphology alone, rather than by searching the literature for specific kinds of objects. In total, the training set consisted of 1400 images: 375 anomalies and 1025 nominal images. We continued building the nominal labelled set during active learning, adding to it substantially to represent further un-interesting morphologies and unaccounted artifacts that were scored highly. The anomalies were primarily mergers (178); galaxies which are undergoing close interactions and coalescing. However, we also uncovered many gravitational lenses that were incorporated into the training set (63). Figure \ref{fig:final-training-set} shows a subset of the training set that was built during the development of $\anomalymatch$.

Increasing the size of the training set and generalising it led to the identification of many different sub-classes of anomalies. However, we were unable to identify new edge-on protoplanetary disks that were not already present in the literature. First, this was due to the known ones becoming a subset of our training set as we add other types of anomalies to it. Secondly, was a limitation of the data we were searching, as observations of edge-on protoplanetary disks in the $F814W$ are rare in the literature.

\section{Methods} \label{methods}
\subsection{\anomalymatch{}}\label{methods:am}
\noindent \anomalymatch{} is an approach combining active learning and SSL that we have developed and optimised for the purpose of identifying anomalies in large datasets. It is fully described, benchmarked, and validated in the companion paper \citet{AMPaper}. We provide a brief description of it here. 

\anomalymatch{} is based on the \texttt{FixMatch} approach \citep{2020arXiv200107685S} as iterated on in \texttt{MSMatch} by \citet{2021arXiv210310368G}. The latter adopted an EfficientNet backbone, support for multispectral data and has demonstrated high accuracy using limited labelled samples through consistency regularisation and pseudo-labelling, effectively handling challenging scenarios such as remote sensing with multispectral data.

Building upon these principles, \anomalymatch{} explicitly frames anomaly detection as a binary classification problem, distinguishing rare anomalies from abundant nominal data, efficiently leveraging a minimal number of labelled anomalies alongside extensive unlabelled data. To effectively address severe class imbalance, the supervised component employs binary cross-entropy loss combined with oversampling of the minority anomaly class, ensuring robust learning from sparse labelled samples.

The unsupervised component applies weak augmentations to generate high-confidence pseudo-labels for unlabelled images, enforcing prediction consistency with strongly augmented versions of these images to exploit intrinsic dataset structure. Crucially, \anomalymatch{} integrates an active learning approach via an interactive user interface (UI), enabling iterative expert validation and labelling of high-confidence anomaly candidates, progressively refining the model and significantly enhancing anomaly detection performance on large datasets with few known anomalies.

This divide between large quantities of unlabelled data and small samples of labelled data is often the case in astronomy. This is especially true for anomalies, where the number of known examples of an object in certain datasets can be less than ten. Therefore, \anomalymatch{} provides a method to expand the catalogues where examples of specific objects are very few. 

\begin{figure*}
    \centering
    \includegraphics[width=0.95\textwidth]{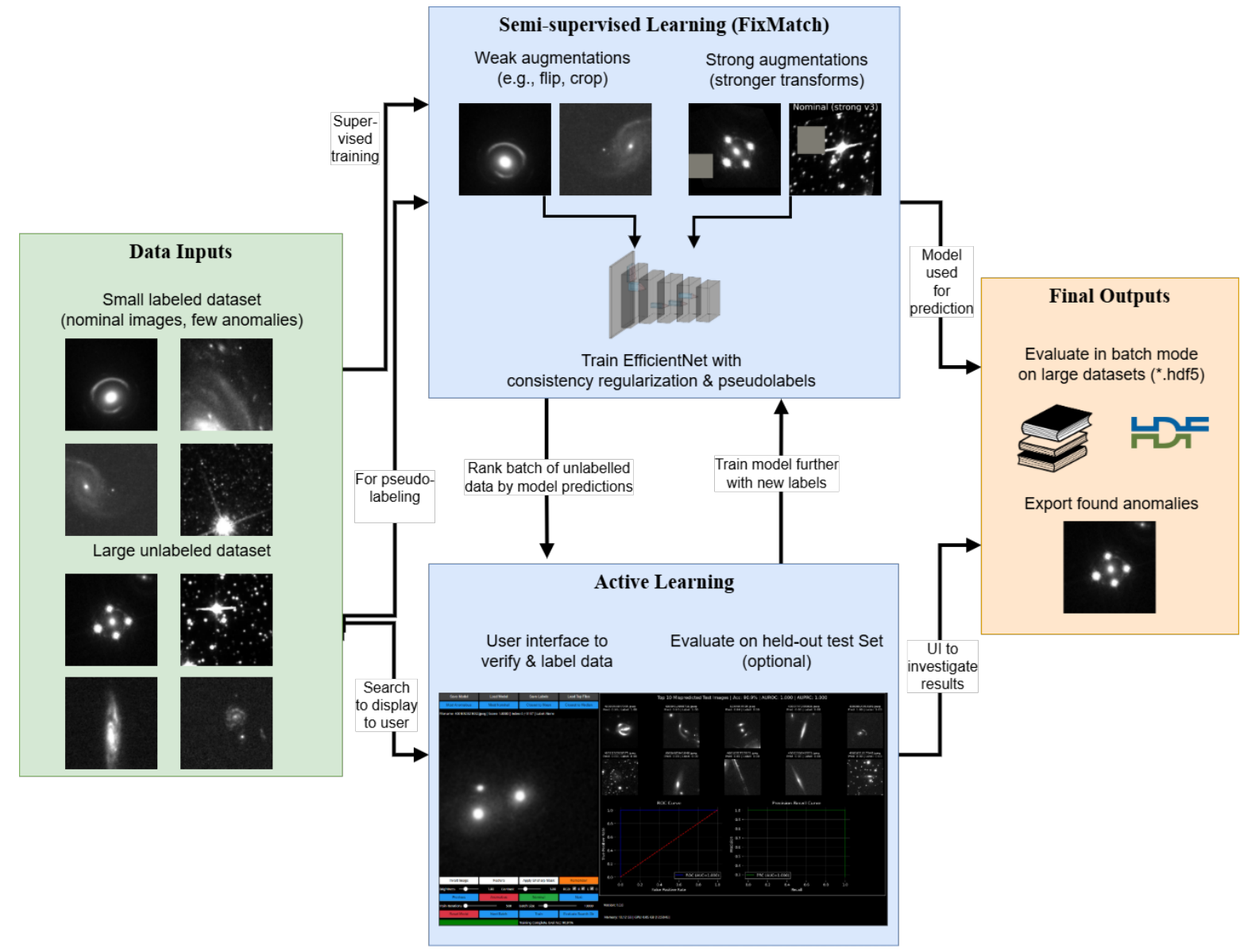}
    \caption{The workflow when using \anomalymatch{}. We leverage both labelled and unlabelled data from a user dataset to train an EfficientNet architecture, and include an active learning loop. Here, the unlabelled data is ranked by anomaly score, the user can extract more examples of the object they are searching for, and add them to their training data. Once the desired model metrics are achieved, the model can be saved and then run across all images in their dataset.}
    \label{fig:workflow}
\end{figure*}

\anomalymatch{} improved on \texttt{MSMatch} by incorporating an active learning loop. The user provides \anomalymatch{} with a set of labelled and unlabelled data, and then conducts an initial round of training. Predictions are then made on the unlabelled data, and an `anomaly score' is given to each image based on the model's classification. The images are ranked by this score, and then shown to the user in an easy to use UI. An example of this is shown in \citet{AMPaper}. The user can then provide additional labels for consecutive training on a dataset that is progressively expanded and refined. This process of active learning is where the the bulk of our initial training set has been collected. Figure \ref{fig:workflow} shows the full \anomalymatch{} workflow. The \anomalymatch{} code will be published on GitHub pending completion of an ESA open-source licensing process.\footnote{\anomalymatch{}: \url{https://github.com/esa/AnomalyMatch}}.

\subsection{Anomaly identification}
\noindent To identify anomalies, the only initial information we have is their morphology. During active learning and inference over the HLA, there was no direct access to further ancillary data. The underlying distribution of anomaly score is highly skewed to nominal images - meaning the model is very confident that an object is not anomalous. Given the strong class imbalance and training on only a limited subset of data, model scores were not calibrated and they are not probabilities. We do not apply a cut to this score but investigate their ranked order.

After training on all labelled data and a subset of the unlabelled data, we visually inspect the 5000 sources with the highest predicted anomaly scores of the entire HLA. Based on either the morphology of the galaxy or the object in the image, we are able to make an initial estimation of the kind of anomaly we find. However, to confirm or reject this, we must turn to the literature. We made use primarily of SIMBAD\footnote{\url{https://simbad.u-strasbg.fr/simbad/}} and the ESASky platform\footnote{\url{https://sky.esa.int/esasky/}}. SIMBAD is an excellent tool to programmatically check if the anomalies we found have any associated papers. We conducted a cone search about the coordinates of each anomaly with a radius of $3"$. We then checked if any source within this radius had was associated to any work at the time of writing. This could then be followed up quickly and efficiently using ESASky.

In many cases, no literature was associated with the anomalies we discovered. For these objects, we determined a classification based solely on the morphology of the system. This classification would be informed by comparing to others we had made where literature was available. Below, we break down our classification scheme for defining different categories of anomalies. Example images for each class are shown in Figure \ref{fig:anomaly-represent} (Section \ref{results}).

\subsubsection{Galaxy mergers}
Galaxy interaction and merging is a well studied phenomenon, with many different algorithms developed for detecting them. Mergers are identifiable as two or more galaxies lying at approximately the same redshift and exhibit signs of a gravitational effect upon one another. Samples are often plagued by incompleteness or contamination from galaxy pairs which are only close together by projection \citep[e.g. discussions in ][]{2018MNRAS.479..415A, 2019A&A...626A..49P, 2024A&A...687A..24M}. Distinguishing between physically interacting galaxy pairs and close pairs is challenging without ancillary data such as velocity or redshift information. Samples of these objects typically include a few thousand systems \citep{2010MNRAS.401.1043D, 2022A&A...661A..52P}.

While there are many stages to a galaxy merger, going from interacting pairs to coalescence, we define any kind of interacting galaxy system as a merger. To identify mergers, we rely on the existence of tidal features between the two systems. These features form through the distortion of the galactic disks as the interaction progresses. This makes our approach primarily sensitive to late-stage mergers, where the systems have already passed each other, resulting in a highly distorted galactic disk with one or more cores visible within it.

\subsubsection{Overlapping galaxies}
Overlapping, or `backlit', galaxies are systems that appear merging in the 2D projection but are at large 3D separations in the plane of the sky. The galaxies show little morphological distortion, with their disks overlapping. This distinguishes them from `close pair' contamination in our merging sample, where the two galaxies are close together in the projection of the sky but not overlapping. Samples of such systems are often small, with the largest being 2,000 from Galaxy Zoo 2 \citep{2013PASP..125....2K}. They are primarily used in the study of dust attenuation independent of dust temperature, and the substructure of galaxies. 

\subsubsection{Gravitational lenses}
Gravitational lenses are the prime example of anomalies in the literature, with many recent works searching for them. They are the product of chance alignment between two galactic systems. The background galaxy's light path is altered in the foreground galaxy's gravitational potential and smeared into an arc around the foreground galaxy. We can observe these gravitational lenses (or, rings if the alignment is complete) around the foreground galaxy, and reconstruct the image of the background galaxy in higher detail than observing it normally. This is due to the lensing having a magnification effect on the background galaxys' light.

The most recent works searching for gravitational lenses used \textit{Euclid} data and have employed both expert labelling \citep{2025A&A...696A.214P} and machine learning techniques \citep{2025arXiv250315330E}. We classify a gravitational lens about an object when we observe these structures around the galaxies in our sample.

\subsubsection{Gravitational arcs}
Gravitational arcs and lenses are results of the same light-bending phenomenon. However, in this work, we distinguish between a strong lens about a galaxy and a strong arc about a galaxy cluster. In our dataset, gravitational arcs are found in source cutouts themselves, rather than closely wrapped around a galaxy within the image. Famous examples of gravitational arcs include the Cosmic Snake \citep{2018NatAs...2...76C} and those used in dedicated galaxy cluster lensing searches \citep{2012ApJS..199...25P, 2019ApJ...884...85C}.

 \subsubsection{Lensed quasars}
 Strongly lensed quasars (often in an Einstein Cross configuration) are the final type of gravitational lensing effect that we classify among our anomalies. These occur when there is the chance alignment between a foreground galaxy and a background highly luminescent quasar, forming distinctive and bright point sources around the foreground galaxy. Very few of these objects are known (<200), with the discovery of a new system often leading to a publication \citep[e.g.][]{2018MNRAS.475.2086A, 2019ApJ...873L..14B, 2023A&A...672L...9T, 2023MNRAS.520.3305L}.

\subsubsection{Jellyfish galaxies}
Jellyfish galaxies are a unique galaxy type found in dense galaxy cluster environments. They are galaxies with high internal gas fractions which are undergoing intense ram pressure stripping (RPS). The dense intracluster medium acts on the internal gas of the galaxy, and strips it into long tendrils in the opposite direction of travel. A distinctive bow-shock forms in the direction of travel of the galaxy, as the gas within is compressed and star formation is likely enhanced \citep{2023MNRAS.524.3502R, 2024ApJ...960...54Z}. Across the literature, samples of these objects rarely exceed 200 systems, although in recent works they are beginning to increase \citep{2017ApJ...844...48P, 2021A&A...650A.111R, 2024A&A...690A.337G, 2025ApJ...982..120F}.

As we classify our anomalies, we find many galaxies which show signs of RPS, i.e. spiral galaxies with large gas reservoirs around them, or long tails with no obvious companion to have created it by interaction or merging. We, therefore, only identify jellyfish galaxies if they show the distinctive bow shock and are located in a cluster environment.

\subsubsection{Clumpy galaxies}
Clumpy galaxies, owing their name to the presence of large and luminous star-forming clumps within them, are gas rich systems which often reside at $1 < z < 3$ \citep{2008A&A...486..741B, 2016ApJ...821...72S}. Samples typically range from 100 - 1000 systems \citep{2015ApJ...800...39G, 2022ApJ...931...16A}. The clumps can easily be mistaken for a second core within the galactic disk leading to possible misclassification as a merger. We ensure that when making either the merger or clumpy classification, we consider for any morphological disturbance to the galactic disk. Also, the existence of more than one secondary core in the galaxy is likely an indication that it is a clumpy galaxy rather than a merger.

\subsubsection{Active galactic nuclei}
Active galactic nuclei (AGN) are the central supermassive black holes of galaxies which are undergoing accretion and growth. They are usually classified from spectral information using diagnostic measurements to identify them based on their emission \citep[e.g., see][]{2003MNRAS.346.1055K}. AGN samples obtained this way are usually large - a few thousand systems. We, however, do not have access to this information and only use the morphology of the system. In rare cases, the AGN can be so luminous that the centre of the galaxy detected as a bright point source. Upon visual inspection, systems hosting a point source at their galactic centre are classified as hosting an AGN.

\subsubsection{Ring and collisional ring galaxies}
We specify two different kinds of ringed galaxies in our anomaly classification scheme: `normal' and collisional ring galaxies. Normal ring galaxies are systems that host a large, empty gap in the galactic disks while a large ring is formed on the outskirts. The formation of such rings is an active area of research. Samples of such rings typically contain several hundred objects \citep{1995ApJS...96...39B, 2014A&A...562A.121C, 2017ApJS..231....2T}.

Collisional rings are formed in galaxy mergers with specific orientations of the impact, when the two galaxies go directly through each other meaning sample sizes are limited. The interaction causes a shockwave to move through the disk incurring a burst of star formation \citep{1996FCPh...16..111A, 2008MNRAS.383.1223M}. We distinguish collisional ring galaxies from ring galaxies as they are far more luminous. They also show disruption to their disks, and the ring itself can be bent or host features. The existence of a secondary galaxy nearby is also accounted for in making this morphology classification.

\subsubsection{Edge-on protoplanetary disks}
As stated in Section \ref{methods}, we originally deployed \anomalymatch{} to search for edge-on protoplanetary disks. These systems are exceptionally rare, with samples not exceeding 25 objects across the literature \citep{2020A&A...642A.164V}. They exhibit a single, dark dust lane across their centres with a distinctive butterfly shape extending in perpendicular directions. They often host a jet, which is also perpendicular to the dust lane. These objects, when known, are very well studied. We, therefore, use this distinctive morphology to identify them in our sample of anomalies. We also closely crossmatch potential candidates with the literature.

\subsubsection{Galaxies hosting a jet}
Galaxies hosting a jet are not uncommon in the literature. Often, in the process of supermassive blackhole growth, long jets will form on either side of the galaxy from the galactic core. While common in the radio or X-ray bands, finding these at optical wavelengths is rare. The most famous example of this is the jet of M87 \citep{1999ApJ...520..621B}, which has been extensively studied with \textit{HST}. 
Optical jets have been identified at various redshifts, typically in samples on the order of tens, and are often used to study AGN properties\cite[e.g.][]{2019ARA&A..57..467B, 2025MNRAS.538.2008K}.

\subsubsection{High redshift galaxies}
During the creation of the images in \citet{2023ApJ...948...40O}, the redshift distribution of the underlying sources was unknown. Therefore, there are many examples in our dataset of sources detected with very low signal-to-noise. In these systems, the morphology of the galaxy can be difficult to distinguish against the background. 

When these low signal-to-noise systems are marked as anomalous by \anomalymatch{}, we broadly label them as high-redshift galaxies, without further classification into anomaly subtypes. Some of these are simply described as `high-redshift galaxies' in the literature as well. 

\subsubsection{Odd galaxies and unknown galaxies}
Finally, we introduce two general classifications that we apply to anomalies that do not easily fit above categories. We label as `Odd' galaxies those which have been scored highly because they host morphological abnormalities or are highly irregular, so they cannot be classified as nominal galaxies, but do not fit our other criteria to be classified as a specific anomaly. Many of the galaxies that meet this definition are located in clusters. In such environments, effects like RPS, harassment and merging lead to major changes in their morphology. However, they do not host a bow shock to be a jellyfish galaxy, and they are not in the process of interacting or merging to be classified as such. These systems also have no available literature to aid in their classification.

The final class is `Unknown' galaxies. These galaxies have a morphology which completely defies classification based on morphology. In many cases, they may not be galaxies. They have no associated literature to aid in their classification. Therefore, we leave these as unknown objects that could potentially serve as new anomalies for further sub-classification, but we are unable to make such a classification in this work.

\subsection{Anomaly score and SHAP distributions}

\begin{figure*}
    \centering
    \includegraphics[width=0.95\textwidth]{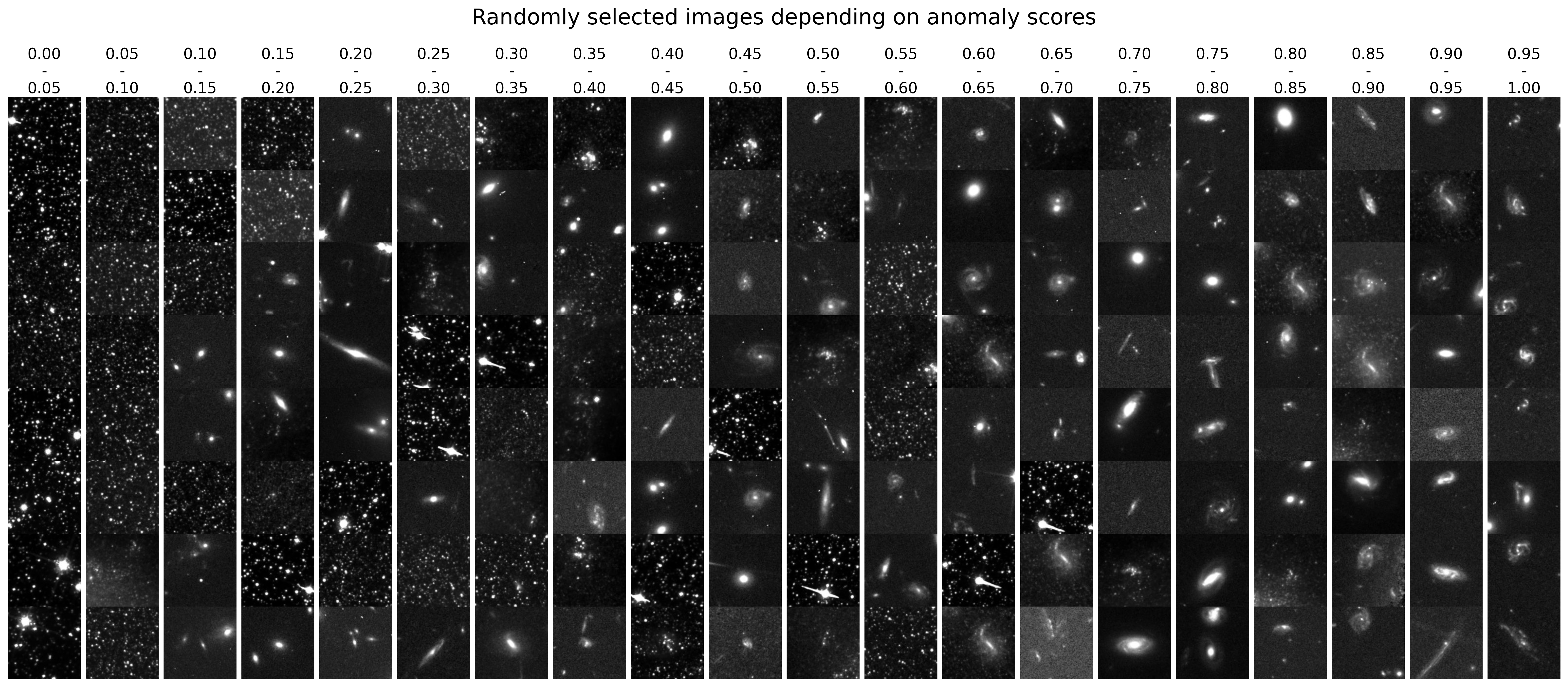}
    \caption{Exemplary anomaly scores on a random subsample of the data. Notably, artefacts are clearly isolated and increasing scores correspond well with increasingly interesting data. The model shows robustness against varying brightness, image noise or differing sizes of objects in the images.}
    \label{fig:score-image-collage}
\end{figure*}

\begin{figure}
    \centering
    \includegraphics[width=0.95\columnwidth]{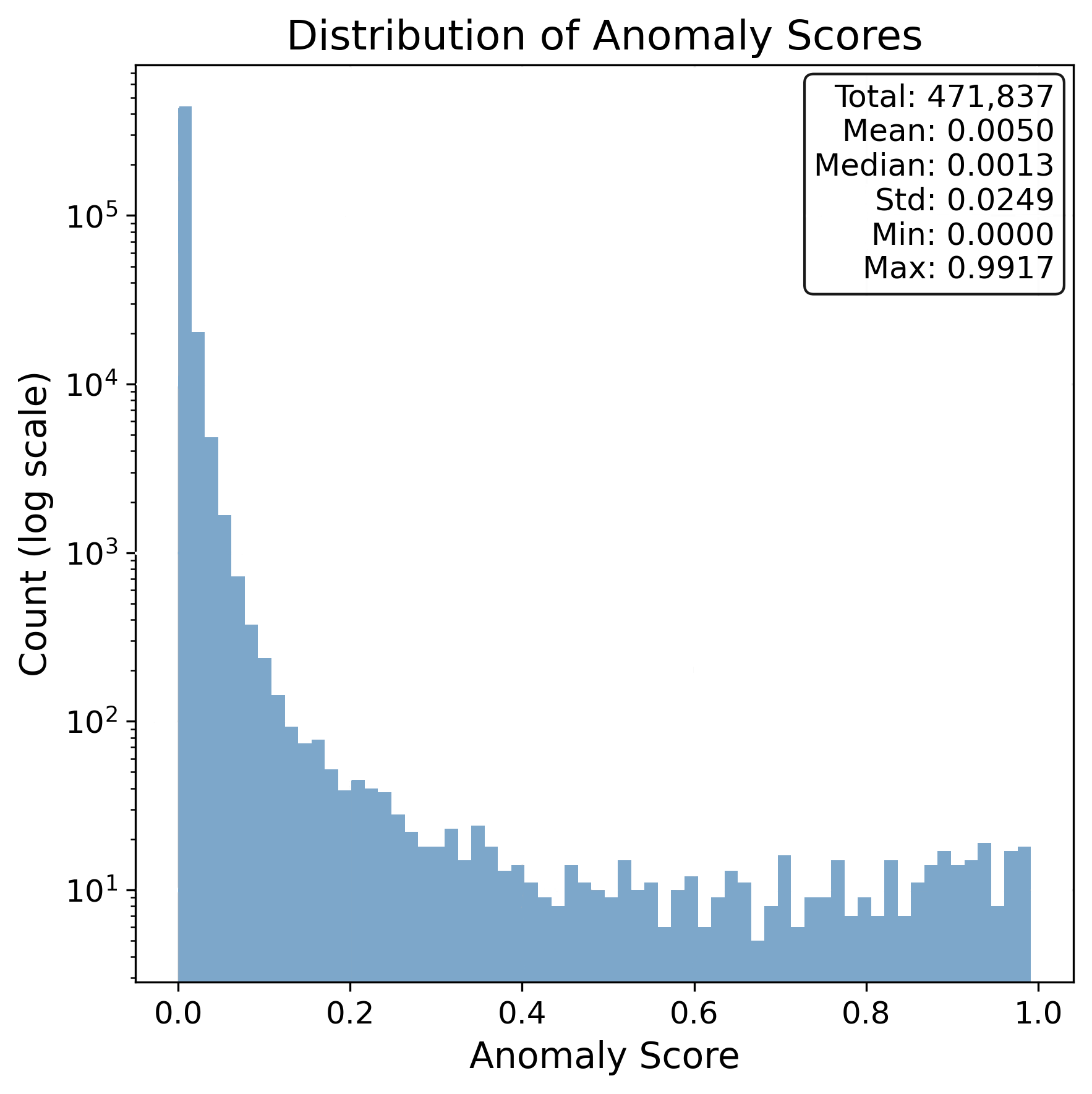}
    \caption{The anomaly score distribution obtained by applying our final trained $\anomalymatch$ model on a random subset of the HLA of $\approx500,000$ cutouts. We find that the distribution is highly weighted to zero, as expected. The majority of our sources are not anomalous.}
    \label{fig:anomaly-score-distribution}
\end{figure}

\begin{figure}
    \centering
    \includegraphics[width=0.65\columnwidth]{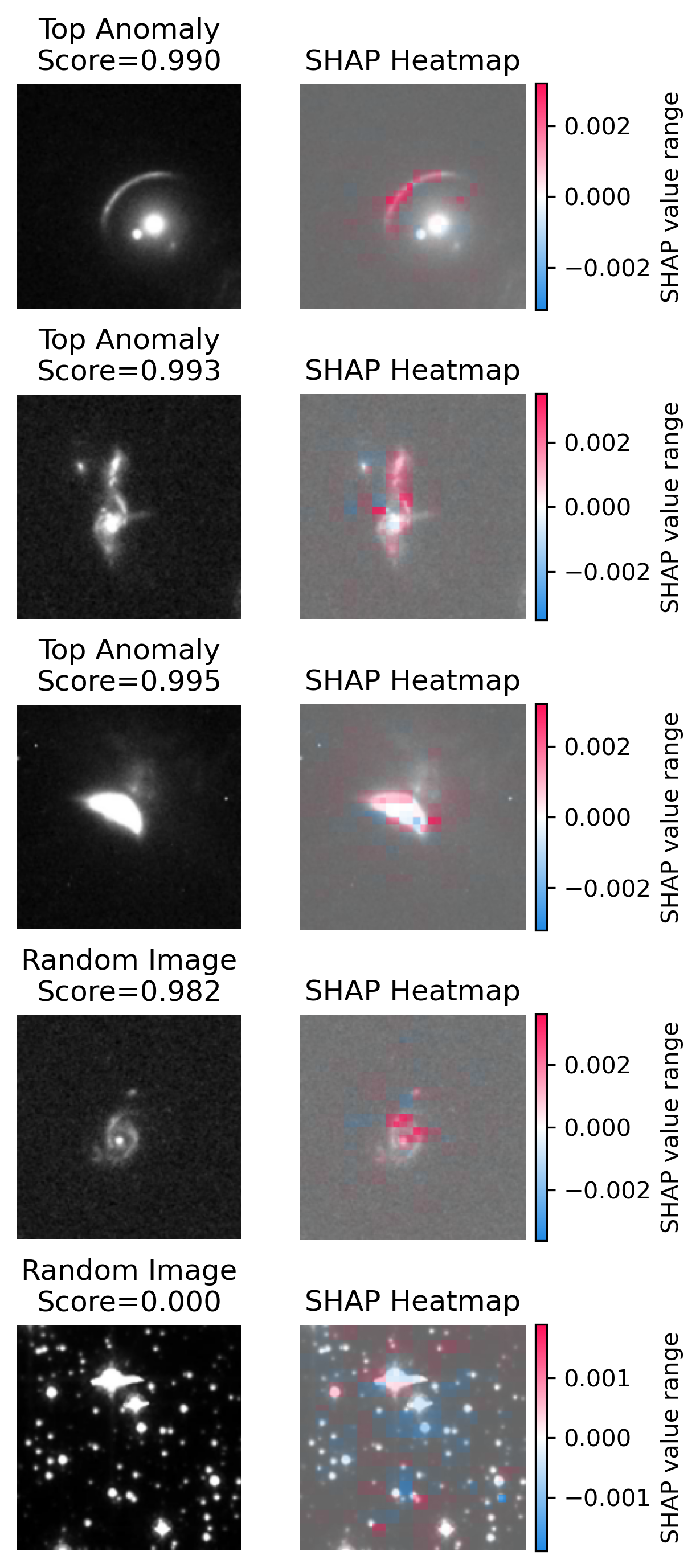}
    \caption{Saliency plots for three highly scored anomalies, one high scoring false positive, and one low-scoring image, presented as heat maps of the SHAP values assigned to each pixel. These visualisations highlight the regions most relevant to the models predictions, often corresponding to features a human might also consider significant (or lack thereof, in the case of the final image). }
    \label{fig:saliency-plots}
\end{figure}

Upon training \anomalymatch{}, we are able to test the model across subsets of our unlabelled data. To demonstrate the performance and output of \anomalymatch{}, we apply it to five random of the stored 1,000 HDF5 files. \anomalymatch{} gives each image a score between 0 and 1 - the anomaly score. Figure \ref{fig:score-image-collage} shows representative examples of different anomaly scores from this distribution. This clearly shows that many of the images in the HLA sample are of star fields, which are ranked very low in anomaly score. As we increase the score, more and more galaxies with ordinary morphologies are revealed.

Figure \ref{fig:anomaly-score-distribution} shows the resultant distribution of anomaly scores across all five HDF5 files ($\approx500,000$ sources). The majority of our sources are given scores $\approx0$. While we do not know the underlying distribution or occurrence rates of anomalies in the dataset, we do not expect many anomalies to appear in a random selection of $0.5\%$ of the data.  At an anomaly score of $\approx1$, we observe anomalies which could be of interest.

To investigate how our model makes its classifications and verify that it behaves as expected, we take examples of the top and bottom scored anomalies and overlay saliency plots on each image. These plots highlight which areas of an image are informing the model to make a classification. We map out changes in the SHapely Additive exPlanations \citep[SHAP][]{NIPS2017_7062} values across each image. This, essentially, shows a weighting of which pixels were relatively more important to the model's classification.

Figure \ref{fig:saliency-plots} shows the results of our saliency mapping. We show three correctly identified anomalies, one false positive and one example of a low scored image. In the correctly identified anomalies, SHAP values are highest at the pixels where the anomaly is within the image. For the lens, this covers the arc itself. For the merger, this is over the tidal features. For the edge-on protoplanetary disk, this is over the butterfly shape of the object. 

The SHAP values for the false positive have an interesting distribution. \anomalymatch{} has highly weighted the core of the galaxy - which could be an AGN in this case - and the extended, loosely wound spiral arm. This has the appearance of a tidal feature in a merger beyond the disk, and towards a second small system which may be a satellite. 

Finally, the last image is of a star field where we can see the SHAP values either weight up or down different areas of the image. However, it has mainly down-weighted the pixels with many stars with empty space between them.

Figure \ref{fig:saliency-plots} demonstrates that \anomalymatch{} has successfully been trained to make anomaly scores based on source(s) within the image, rather than on another property of the image. Even in images containing only one source, it weights up pixels based on dividual sections of the galaxy rather than the entire object in the image. This is shown with the lens image (image 1), where $\anomalymatch$ primarily uses the lens itself to make the classification instead of the galaxy within the image. 

\subsection{ESA Datalabs}
\noindent For storing data and conducting training and inference over the images, we utilise the platform ESA Datalabs. This platform allows for direct access to data from many different of ESA's observatories and missions where it is a partner with other agencies, removing the requirement for downloading of observations or movement of data. We take advantage of the newly integrated GPU cluster which is available via the platform, and allows us to both train the model and conduct inference efficiently, as proved by the timescale of this project. 

Fully training the model over our training set (1,400 labelled, and $\approx99,000$ unlabelled images) took less than four hours on a GPU via ESA Datalabs. Making classifications over all $\approx100$ million images took 2.5 days by running the code on ESA Datalabs with no user input, or intervention. The main bottleneck in efficiency was loading the images into memory to then be inferred over. This was done via HDF5 files in batches of 100k.

This high level of efficiency shows that ESA Datalabs is an ideal platform for large-scale data exploration. With the addition of GPUs to the platform, applying machine learning algorithms to large quantities of observational data is straightforward. This will facilitate further exploration of a wide range of observational datasets, including JWST, \textit{Euclid} and \textit{HST}.

\section{Results}\label{results}
\noindent To perform the anomaly sub-classification, we initially selected the top 5000 scored anomalies from the HLA. First, we apply de-duplication to these 5000 samples. The output from \anomalymatch{} is a CSV file containing the filename of the source and the score given to the source. In our case, the source filename is the source ID found in the HSC. We cross match each of our sources with the HSC based on this, and extract their coordinates. We then apply an aggressive radial cross match between each of the sources within $10"$.

We apply such a large cut as the likelihood that two of our anomalies are within this separation is low. It allows us to completely de-duplicate the final sample, and ensure we are not using expert time re-inspecting duplicate entries. This reduces the number of anomalies from 5,000 to 1,338 unique images. This also highlights the high level of shredding and duplication in the HSC. Upon de-duplication, DOR acted as the expert classifier and classified each anomaly according to the system defined in Section \ref{methods}.

\subsection{Detected anomalies}
\begin{figure*}
    \centering
    \includegraphics[width=0.65\textwidth]{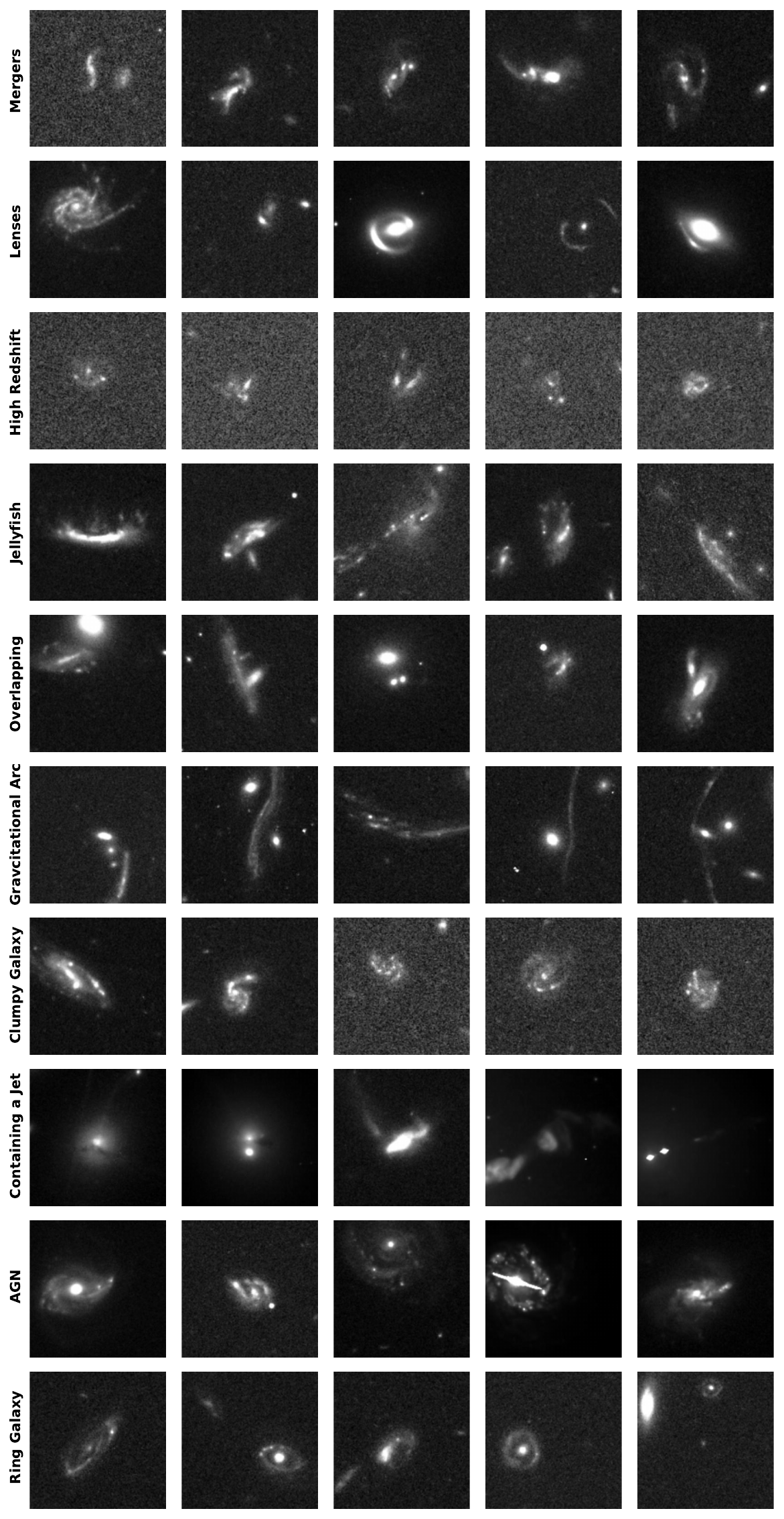}
    \caption{Five examples of every anomaly sub-class for which we found at least five objects (excluding lensed quasars for later discussion), selected as representative of that sub-class.}
    \label{fig:anomaly-represent}
\end{figure*}

The unique 1,338 anomalies taken from the ranking of highest anomaly score were classified into their sub-classifications. This was done by a combination of accounting for morphology and searching the literature for works related to each anomaly. Table \ref{tab:anomalies} shows a breakdown of these classification of the anomalies. Figure \ref{fig:anomaly-represent} shows a representative sample of all sub-classifications where we found at least five anomalies, except for our `odd', `nominal' and `unknown' classifications, which will be presented later in the section. The `nominal' classification is our false positive rate, where $\anomalymatch$ returned a normal source.

\begin{table}
  \caption{Breakdown of anomalies found in the development of the $\anomalymatch$ algorithm.}
  \label{tab:anomalies}
      \begin{tabular}{ccc}
        \hline
        Classification & N Found & N Referenced \\
        \hline\hline
        Merging Galaxy & 629 & 212 \\
        Gravitational Lenses & 140 & 54 \\
        Odd Galaxy & 164 & 60 \\
        Nominal Galaxy & 163 & 60 \\
        Unknown Morphology & 43 & 0 \\
        High Redshift Galaxy & 28 & 7 \\
        Jellyfish Galaxy & 37 & 18 \\
        Overlapping Galaxy & 39 & 26 \\
        Gravitational Arc & 39 & 19 \\
        Clumpy Galaxy & 11 & 5 \\
        Galaxy Hosting a Jet & 13 & 5 \\
        Galaxy Hosting AGN & 8 & 4 \\
        Ring Galaxy & 12 & 5 \\
        Lensed Quasar & 5 & 5 \\
        Collisional Ring Galaxy & 2 & 0 \\
        Edge-On Protoplanetary Disk & 2 & 2 \\
        Galaxy Hosting Supernova & 2 & 2 \\
        Submillimetre Galaxy & 1 & 1 \\
        \hline
      \end{tabular}
\end{table}

The largest population of anomalies we find are merging or interacting galaxies ($\approx$50\%). These are likely the most common type of anomaly that we are searching for, as well as the most distinctive. With our aggressive de-duplication, these represent 629 merging systems rather than individual systems. The cutouts used in \anomalymatch{} contain a field of view of $7.5" \times 7.5"$, therefore galaxy mergers contained in the same source cutout will be de-duplicated under one ID. Following our definition, not all of our classified merger anomalies have secondaries. Many are highly disturbed merging or post-merger systems, with only one system in the image. 

It is also important to note that for some of merger classifications, the full extent of the system is not visible in the field of view of the cutout. However, the tidal features of the system are and therefore, the anomaly score was ranked highly. Looking at these systems within ESASky reveals the full system and associated literature when the field of view is adjusted.

The second most numerous anomaly we detect is gravitational lenses (and arcs, if classified under the same category). These are currently the main anomaly being searched for in the field, and therefore, made up a large portion of the training set. Due to the shape and position of lenses, the main source of contamination in our results will be spiral arms, or other features which appear like arcs around galaxies. The different luminosity of the lens aids in the classification, however this is not always the case. We identify many gravitational lenses that are already identified in the literature - but many candidate new lenses. As we do not attempt to model, and confirm them, in this work we point out that these are only candidate lenses which require either follow-up observations or modelling to confirm.

We do not include our sample of 39 gravitational arcs from our lens classification here. Often, these arcs - formed by clusters rather than galaxies - are so large, that they extend out of the field of view of our source cutouts. We see the true extent of these systems in ESASky, although only parts of them are flagged as anomalies here. Other cutouts containing the arc are removed in our aggressive cross matching and de-duplication of $10"$. We also ensure that each of our 39 arcs is unique using ESASky. 

We find a population of what we have termed `high redshift' galaxies ($z > 1$). These are systems that appear highly disturbed, or affected in the image, but are at the threshold of being detected. Looking at those which are referenced, we find that these systems are representative of high redshift galaxies. We also find them to be small and clumpy.

The clump classification we make provides a sample of 11 systems. These galaxies could also be mistaken for merging galaxies, as they often look irregular compared to the general galaxy population. However, the disks often host many more than one other core - which would be expected in mergers - and often these are contained in spiral arms, which would be destroyed in the merging process.

We find 37 different jellyfish galaxies in our 1,338 unique systems. These have been classified as such as they clearly have stripping occurring, and they are residing in a dense cluster environment. On each of the examples shown in Figure \ref{fig:anomaly-represent}, the bow shock is present in the direction of travel. 

Next, we identify a similar number of overlapping galaxy systems, with the whole system included in the field of view. A very small sample (13) of our found anomalies are galaxies which host relativistic jets. These can rarely be detected in the $F814W$ filter of \textit{HST} (the most famous being M87, which is shredded and appears multiple times in the HSC). Our systems are much smaller and often show a small jet or object moving away from the galactic core. This classification was made by comparing to the literature, where galaxies with active jets made up just two of the samples we found. However, upon recognising this morphology, we were able to make more classifications.

Our final two classifications are AGN-host galaxies (8) and ring galaxies (12). Our AGN classification is primarily from the literature, where all but one of these systems has an associated reference. Otherwise, the requirement for ancillary data would make this classification difficult with \anomalymatch{} alone. Finally, we identify ring galaxies by their morphology as the structure is easily recognisable compared to others.

\begin{figure}
    \centering
    \includegraphics[width=0.85\columnwidth]{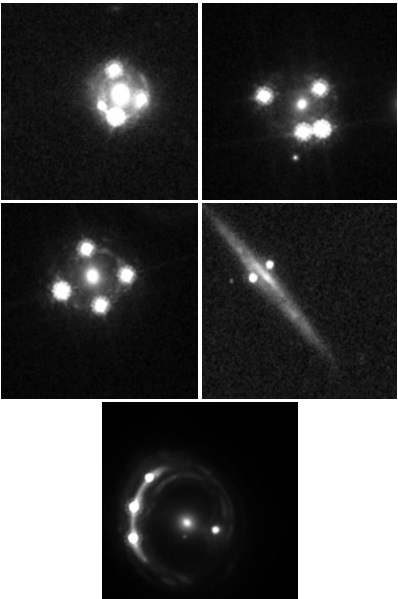}
    \caption{The five lensed quasars (Einstein crosses) found in this work. All of which are represented in the literature. However, using \textit{HST}, we would not expect to find unreferenced systems.}
    \label{fig:einstein_crosses}
\end{figure}

\anomalymatch{} successfully detected several types of anomalies for which we had provided no explicit training examples, such as lensed quasars. While these unseen anomaly types likely share morphological features with our trained classes (e.g., lensed quasars may exhibit arc-like features similar to gravitational lenses or multiple bright regions like mergers), their detection demonstrates the method's ability to generalise to morphologically similar but distinct anomaly types. However, we acknowledge that unseen anomalies with fundamentally different morphologies from our training set would likely be missed. Figure \ref{fig:einstein_crosses} shows the four systems that we found in HLA. We initially provided \anomalymatch{} with no examples of these, however, even in the first training iterations, it identified these anomalies and scored them highly. Therefore, we added one to the training set and identified four additional lensed quasars.

\begin{figure}
    \centering
    \includegraphics[width=0.95\columnwidth]{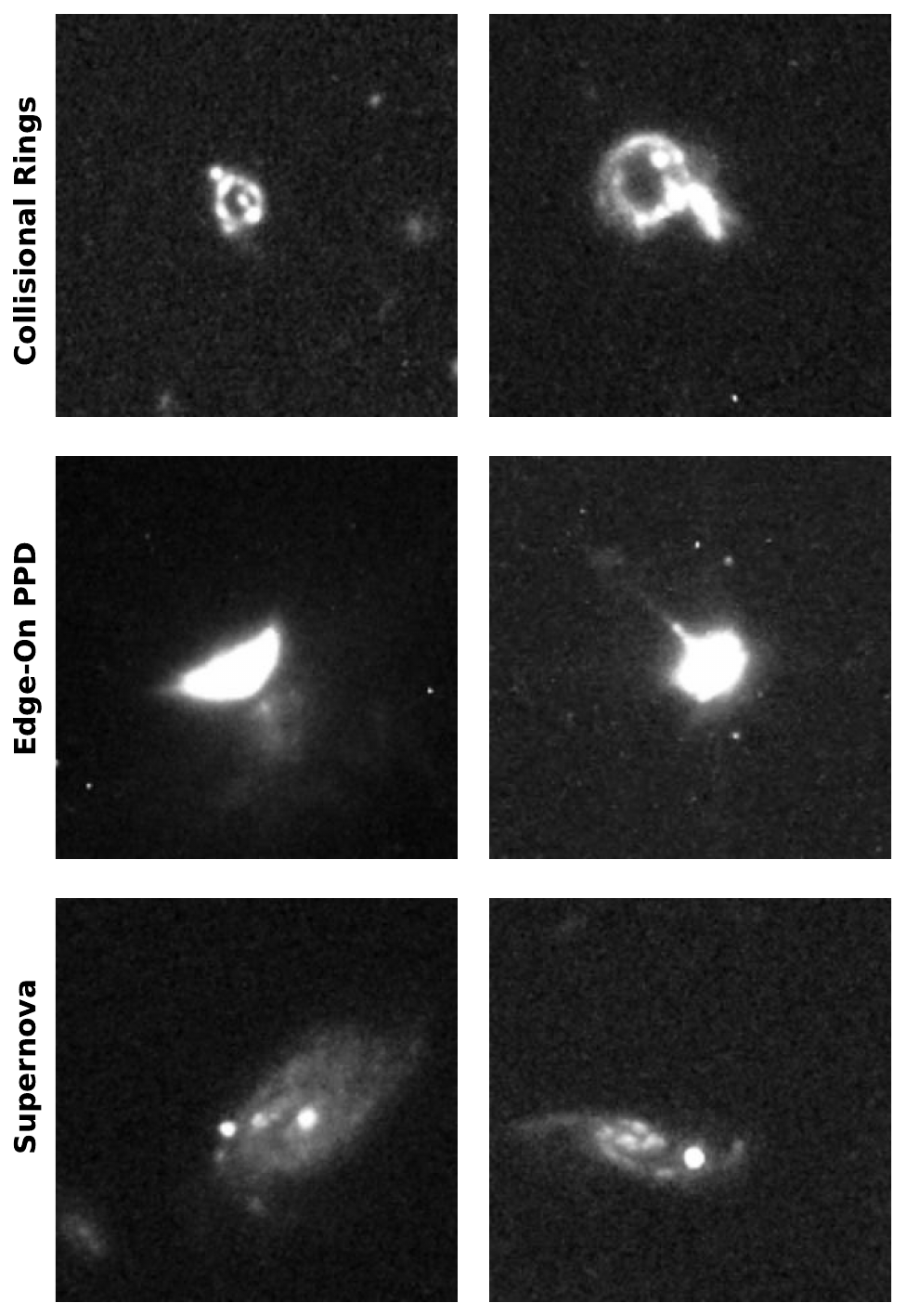}
    \caption{The smaller anomaly samples found in this work. We find two collisional ring systems, two further edge-on protoplanetary disks and two galaxies containing supernovae.}
    \label{fig:small-samples}
\end{figure}

Figure \ref{fig:small-samples} shows the identified collisional rings, edge-on protoplanetary disks and galaxies hosting a supernova. While ancillary data are required for the classifications of supernova hosts and edge-on protoplanetary disks, we confirm these with the literature here. This is also true for our classification of a submillimetre galaxy, where the object we identify has a large repository of literature associated with it. However, our collisional rings are not represented in the literature, and are therefore, classifications we have made by checking examples of known collisional rings, and confirm their morphology is similar. 

\begin{figure*}
    \centering
    \includegraphics[width=0.80\textwidth]{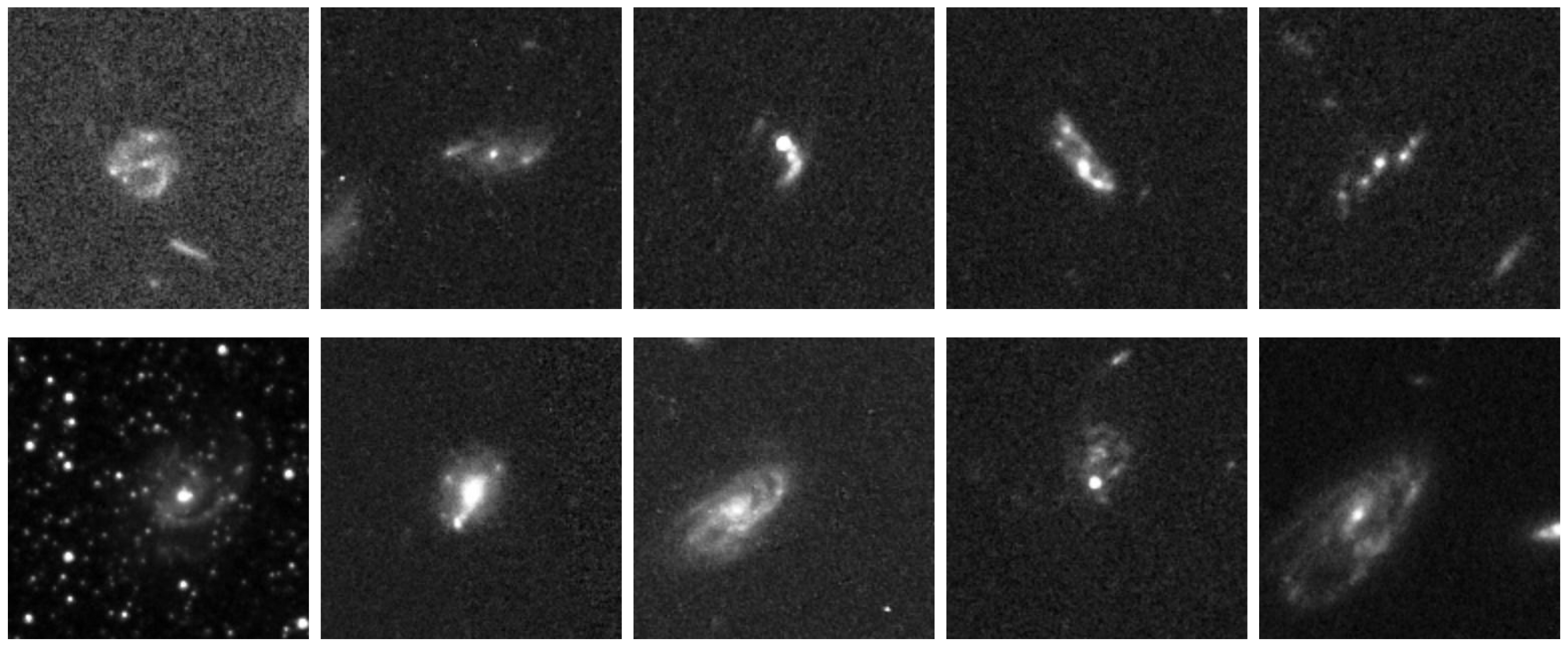}
    \caption{Ten examples of sources to which \anomalymatch{} gave a high anomaly score, and we have classified as odd morphologies. These galaxies are close to being classified as nominal galaxies - i.e. of little interest to the community - but have odd morphological features. These are often due to these galaxies residing in dense environments.}
    \label{fig:odd-ims}
\end{figure*}

Finally, we include three general classifications in our released catalogues: odd, nominal and unknown galaxy morphologies. As stated in our classification scheme, we define odd galaxies as those which show an odd morphology which could be identified as due to interaction or RPS. However, we find that they do not fit this criteria. Often, they are galaxies within a dense environment showing some distortion. Figure \ref{fig:odd-ims} shows ten representative examples of this distortion possibly due to harassment by other cluster members. Some of these odd galaxies are not in a cluster, and simply are irregular galaxies that \anomalymatch{} has scored highly. It is not unreasonable for these to have been detected with our extended training set of different anomalies, particularly if including mergers.

\begin{figure*}
    \centering
    \includegraphics[width=0.80\textwidth]{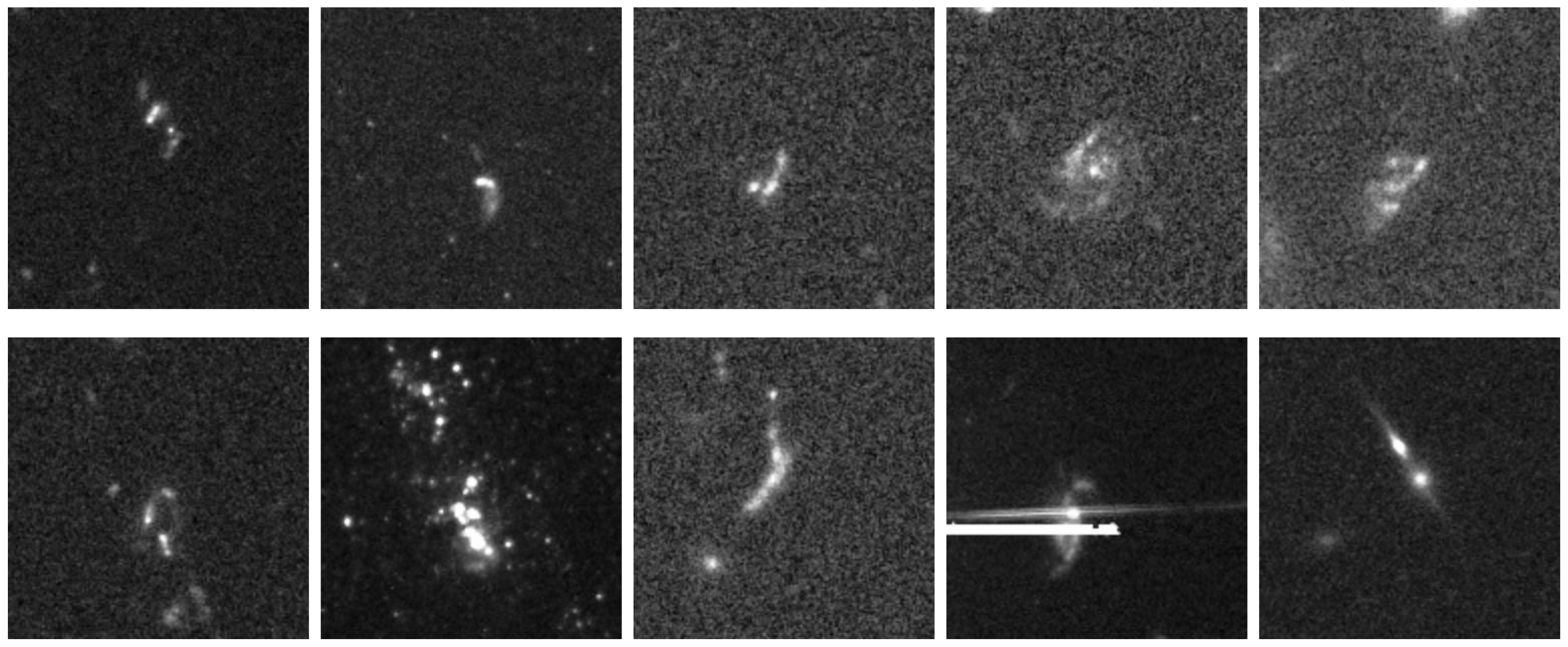}
    \caption{Ten examples of sources to which \anomalymatch{} gave a high anomaly score, but we have visually classified as nominal images. These include sources which exhibit some features, but are too small in the field of view to give definitive classification. It also includes some star fields, as well as artifacts in unique alignments.}
    \label{fig:nominal-ims}
\end{figure*}

\begin{figure*}
    \centering
    \includegraphics[width=0.80\textwidth]{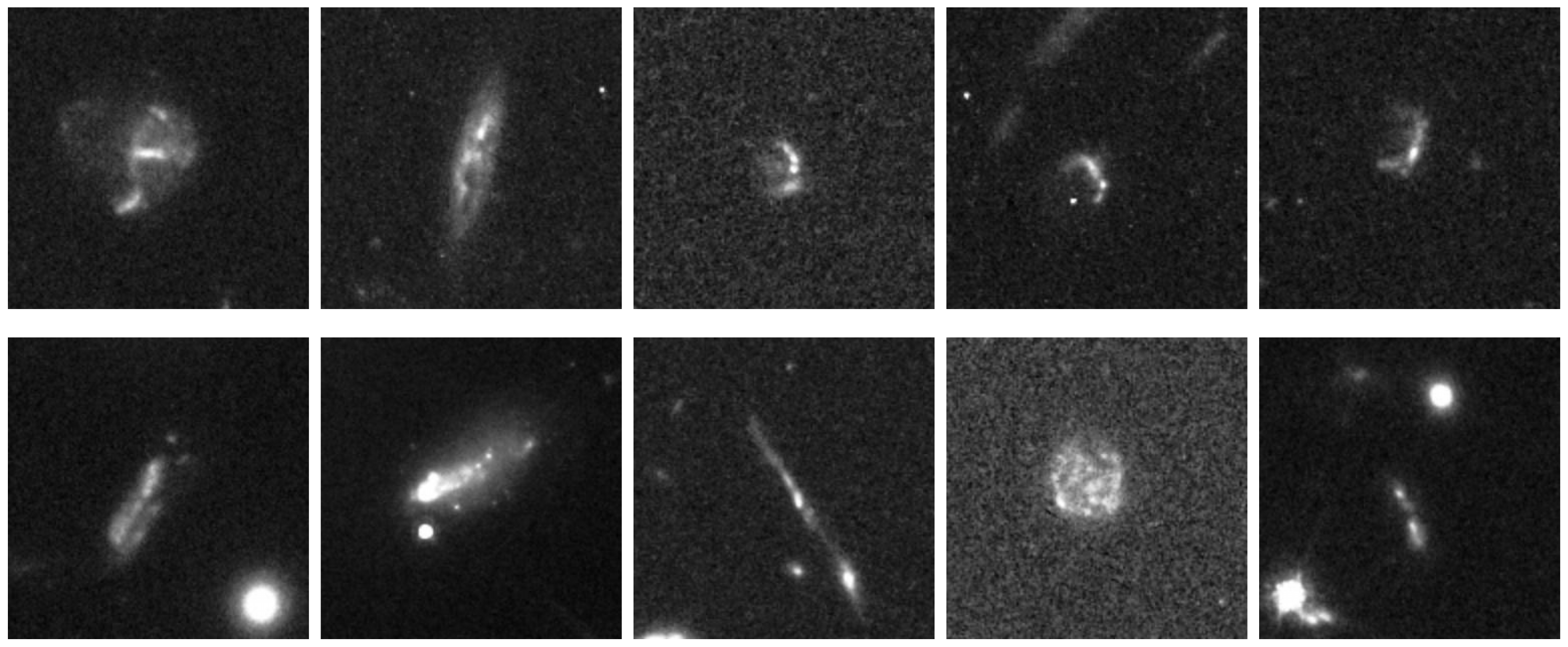}
    \caption{Ten examples of sources to which \anomalymatch{} gave a high anomaly score, but we have been unable to morphologically classify. Many of these could be jellyfish galaxies, as they appear to show RPS signatures. However, they do not reside in a dense cluster environment. Others show curved morphologies, which are do not fall into our classification scheme here. }
    \label{fig:unknown-ims}
\end{figure*}

Galaxies we classify as nominal by visual inspection, but that received a high score by \anomalymatch{}, represent the contamination in our final anomaly sample. We find a contamination rate of $\approx10\%$. Figure \ref{fig:nominal-ims} shows a representative example of ten sources. Primarily, these sources are very small sources which exhibit some potential merging behaviour, but it is difficult to make a specific classification. Other examples include sources with artifacts from the observation itself, aligned with a point source and star fields. Star fields with this morphology are often either from the Panchromatic \textit{Hubble} Andromeda Treasury \citep{2012ApJS..200...18D} or studies of large globular clusters with the ACS instrument. Therefore, many of the nominal images we find here are anomalous when compared to the rest of the data, but they are of limited value to the astrophysical community. 

Finally, we find 43 objects with morphologies defying classification. Some of these objects may not be galaxies but rather other objects that we have limited expertise in classifying. Figure \ref{fig:unknown-ims} shows ten example sources we have been unable to classify. We release these to the community for further discussion, or use, but do not attempt to make morphology classifications here. None of these objects have definitions in the literature.

We release each of these catalogues as a machine-readable table (MRT). The table contains the source identifier from the HSC, its right ascension and declination, our classification and whether the source has been classified by using the literature. An example of the first five rows of the table are shown in Table \ref{tab:data-release}.

\begin{table*}
    \caption{First five rows of the MRT released with this work.}
    \label{tab:data-release}
    \centering
    \begin{tabular}{ccccc}
        \hline
         SourceID & RA & Dec & Classification & Referenced \\ 
         \hline\hline
          4000862191779 & 8.461384 & -7.833325 & Lens & False  \\ 
          4000705851114 & 186.877395 & 33.544637 & Lens & False  \\
          6000532112741 & 342.211252 & -44.513375 & Merger & True  \\
          4000905661818 & 34.404996 & -5.224907 & Lens & True  \\
          6000265754287 & 53.175990 & -27.773749 & Merger & True  \\
         \hline 
    \end{tabular}
    \tablefoot{The table contains the (1) Source ID from the HSC, (2) Right Ascension in degrees, (3) Declination in Degrees, (4) classification in this work and (5) whether the classification was made using the literature.}
\end{table*}

For the ease of use of our catalogue, we also release all of the sources and table on Zenodo (DOI: \href{https://zenodo.org/records/15401617}{10.5281/zenodo.15298641}). Here, the images are contained in separate .zip files based on their anomaly classification. The filenames are the Source IDs of the sources.

\section{Discussion}\label{discussion}
\noindent Conducting robust and complete anomaly detection is non-trivial. As stated previously, utilising supervised machine learning methods suffer from a lack of training data, and the complex and variable nature of anomalies to be found. Using this test run of the neural network $\anomalymatch$ during development, we have been able to extract a range of anomalous objects from a limited training set. With this SSL approach, $\anomalymatch$ is able to distinguish unique morphologies from general ones using an incredibly limited set of labels.

Other works focused on anomaly detection often must begin from using similarity searches, to citizen scientists to identify large samples for further searches \citep{2008MNRAS.389.1179L, 2021A&C....3600481L, mantha20224, 2025arXiv250315324E}. Conversely, \anomalymatch{} can be used as a baseline approach to grow existing samples of anomalies from the literature. It can also be employed in a flexible manner, where other anomalies not included in the initial labelled dataset can be detected and extracted during the hunt for other objects.

This is the main difference between the envisioned implementation of \anomalymatch{} and what we find here. We started by searching for edge-on protoplanetary disks - as they are very rare objects - but expanded our training set to include other objects like mergers, lenses and jellyfish galaxies upon their serendipitous discovery. This approach results in a varied output which will contain objects a user is not interested in, but the inclusion of the active learning loop in the training process allows them to filter out unwanted anomalies and retain those which are desired. For instance, in this work, the majority of the anomalies we find are mergers. Mergers are reasonably common - although difficult to build large samples of. In the future, we would use the active learning loop to remove these objects, increasing purity and accuracy in finding a desired anomaly.

In this work, we find 810 unreferenced objects in our sample of 1,338 objects. However, by the nature of using the HLA and, specifically, \textit{HST} to search for anomalies, we are limiting our capability for identifying new objects. The \textit{HST} is not a survey telescope: observations must be applied for, and PIs specifically request the coordinates and objects they wish to observe. Therefore, any anomalies we do find are either in the background of observations of another astrophysical source of interest, or are the target of the observation in question. The latter is particularly true for edge-on proto-planetary disks, where these targets of exceptional interest are rarely observed `by accident' in the background of targeted observations. The fact that $\approx65\%$ of our anomalies are not represented in the literature thus shows the exceptional potential of $\anomalymatch$ for application to survey telescopes' data.

With survey telescopes, much of the data will be completely unexplored and rarely looked at besides with machine learning algorithms. This role is where \anomalymatch{} will excel and, as shown in this work, be able to identify objects of various morphologies that can then be classified into different categories. Missions such as \textit{Euclid} and Vera C. Rubin observatories are perfect for this role, where Terabytes of data will be collected and processed per night. 

In the future, \anomalymatch{} will be used to search for specific types of anomalies that the user can hone in on. By ranking objects in this way, we will be able to rapidly find and increase the size of our samples of different anomalies while using visual inspection. As found in \citet{AMPaper}, anomalies identified with the 1\% highest scores have a high precision, i.e. chance to be only or almost only anomalies - assuming such objects are present in the data of them in the first place. However, extracting them still requires visual inspection by the user.

\section{Conclusion}\label{conclusion}
\noindent We have developed and deployed an innovative approach based on semi-supervised and active learning for the purpose of identifying sources with anomalous morphologies. The method - \anomalymatch{} - has been created as an out-of-the-box tool, which has been integrated into the ESA Datalabs platform. It has an intuitive and interactive UI, which seamlessly combines semi-supervised learning methods with active learning so a user can identify objects of interest efficiently, and with few labels. The full benchmarking and testing of this algorithm is described in \citet{AMPaper}.

In this work, we release the anomalies that we discovered in the HLA while developing this approach. Using a large test set of 99.6 million image cutouts in the $F814W$ filter of \textit{HST}, we ranked them all by anomaly score. To start, we aimed to identify more edge-on protoplanetary disks. Using a training set of just three labelled objects, and a large sample of nominal and unlabelled data, we searched the archive for objects of interest. During active learning, we identified many other objects of astrophysical interest and grew the training set based on what we found.

Upon completion of development, and training this model and applying it across the HLA, we selected the top 1,338 de-duplicated sources. We then classified these into different sub-classes of anomalies based on morphology. These included 629 mergers, 140 gravitational lenses, 37 jellyfish galaxies and 2 edge-on protoplanetary disks. Using SIMBAD and ESASky, we conducted a literature search based on the coordinates of each object, and to investigate if they had been included in other catalogues and samples. We find that  $\approx65\%$, 811, of the objects do not appear in the literature.

We release a machine readable table containing source IDs, positions and classifications of all sources we have discussed in this work. We also release both the tables and images of each object on Zenodo (DOI: \href{https://zenodo.org/records/15298642}{10.5281/zenodo.15298642}), where these samples are available for download.

The large number of unreferenced sources is excellent, and shows \anomalymatch{}'s capability for identifying new systems. This is especially notable given that our data comes from \textit{HST}, where each observation is targeted - meaning that either a PI requested to observe the anomaly we have detected, or it is a previously unnoticed background system.

This shows the potential of \anomalymatch{} for use in large surveys, such as \textit{Euclid} or Vera C. Rubin Observatory, which will observe large areas of the sky with no knowledge of the objects contained within. The large volume of data that will be produced per night will be impossible to manually search and visually inspect. Therefore, using algorithms where small amounts of data are required for training will be paramount for anomaly detection in these large surveys.

\anomalymatch{} thus fills a particularly fruitful niche. We will be able to use this method to gradually grow our small samples of different anomalies, while also uncovering anomalies with completely new morphologies. 

\anomalymatch{} is freely available on the ESA Datalabs platform with an open-source release pending a successful licensing process currently underway at ESA. The software will be available on GitHub\footnote{\url{https://github.com/esa/AnomalyMatch}}. It has been developed with GPU capabilities to be able to seamlessly explore ESA data archives and search them for anomalies of interest to other users. This powerful addition and implementation of SSL will enhance our ability to detect anomalies across the field.

\begin{acknowledgements}
\noindent The authors would like to acknowledge and thank Chris Evans, Sandor Kruk, Maria Teresa Nardone, Pedro Mas Buitrago, and Laslo Ruhberg for their valuable feedback on this manuscript. DOR acknowledges the support of the ESA Research Fellowship in Space Science program. This work was published as part of his research fellowship. The authors also acknowledge and thank the thorough review of this work by the external referee. Their reports strengthened this work significantly.

This work makes extensive use of datasets from the \textit{Hubble} Legacy Archive. Stored here are observations with the NASA/ESA \textit{Hubble Space Telescope}, which is a collaboration between the Space Telescope Science Institute (STScI / NASA), The Space Telescope European Coordinating Facility (ST-ECF/ESA) and the Canadian Astronomy Data Centre (CADC/NRC/CSA). This work also made extensive use of the\textit{Hubble} Source Cataloue, derived from the HLA. This is fully described in \citet{2016AJ....151..134W}. To efficiently interface with the HLA, ESA's science platform ESA Datalabs was used. ESA Datalabs (data.esa.int) is an initiative by ESA's Data Science and Archives division in the Science and Operations Department, Directorate of Science. \citet{2024sdm..book....1N} describes the platform.

To study the objects we found, and aid in their classifications, extensive use of the ESASky platform and SIMBAD database were used. The ESASky platform was developed by the ESAC Science Data Centre (ESDC) team and maintained alongside other ESA science mission's archives at ESA's European Space Astronomy Centre (ESAC, Madrid, Spain). A full description of the platform can be found in \citet{2017PASP..129b8001B} and \citet{2018A&C....24...97G}. This research has also made use of the SIMBAD database, operated at CDS, Strasbourg France. It is fully described in \citet{2000A&AS..143....9W}.

This research made use of many open-source Python packages and scientific computing systems. These included \texttt{Matplotlib} \cite{matplotlib}, \texttt{Pandas} \citep{pandas},and \texttt{numpy} \citep{numpy}. This work also extensively used the community-driven Python package \texttt{Astropy} \citep{astropy:2018, astropy:2022}.

This work made extensive use of the new tool \texttt{AnomalyMatch}, described in the companion paper \citet{AMPaper}. For writing and editing, no additional AI tools were used in this manuscript. 
\end{acknowledgements}

\bibliographystyle{aa}
\bibliography{references}

\label{lastpage}

\end{document}